\documentclass[aip,graphicx]{revtex4-1}
\usepackage{graphicx}
\usepackage{dcolumn}
\usepackage{bm}
\usepackage{caption}
\usepackage[utf8]{inputenc}
\usepackage[T1]{fontenc}
\usepackage{mathptmx}
\usepackage{hyperref}
\hypersetup{hypertex=true,colorlinks=true,linkcolor=blue,anchorcolor=blue,citecolor=black}
\usepackage{etoolbox}
\usepackage{array}
\usepackage{lineno}
\usepackage{setspace}
\usepackage{float}
\usepackage{amssymb,amsmath,bm,booktabs}
\usepackage{makecell}
\usepackage[lined,boxed,commentsnumbered]{algorithm2e}
\draft 
\makeatletter
\def\@email#1#2{%
	\endgroup
	\patchcmd{\titleblock@produce}
	{\frontmatter@RRAPformat}
	{\frontmatter@RRAPformat{\produce@RRAP{*#1\href{mailto:#2}{#2}}}\frontmatter@RRAPformat}
	{}{}
}%
\makeatother
\begin{document}


\title[Unsteady DPM Incipient Around Monopile]{Simulation of sediment incipience around the vibrating monopile}



\author{Yuxuan Shi}
\email{20004010053@stu.csust.edu.cn}
\altaffiliation[Also at ]{School of Hydraulic and Environmental Engineering, Changsha University of Science and Technology.}
\author{Yongzhou Cheng}%
\email{chengyz@csust.edu.cn}
\affiliation{ 
	Key Laboratory of Water-Sediment Sciences and Water Disaster Prevention of Hunan Province
}%

\date{\today}

\begin{abstract}
Based on the fluid volume fraction-large eddy simulation-discrete phase coupling model, the water and sediment flow around the vibrating monopile was simulated. The discrete phase momentum model was enhanced using the force distribution of the particle phase on the bed surface. Additionally, the flow dynamics and sediment incipience around a vibrating monopile was replicated through the use of a sediment incipience Shields number. The results suggest that monopile vibrations amplify the negative incipience of sediment in both the front deceleration region and the wake reflux region. Furthermore, within the periodic flow field surrounding the monopile, the temporal changes in the sediment’s incipience Shields number does not consistently match the temporal changes in the spatial average velocity. 
\end{abstract}


\maketitle 

\section{Introduction}

With the continuous development and application of new energy technology, research on offshore monopile wind turbines has received widespread attention. In such engineering structures, the application of monopile structures is very common, and the phenomenon of flow around blunt body is very common\cite{ref-1,ref-2}. Due to the frequent installation of such structures in offshore areas, the phenomenon of water carrying sediment and sediment movement influenced by water flow is common in the flow around monopile. In addition, the monopile affects the shape of the convection field, making the development process of water sediment two-phase flow around the pile more complex. This water sediment two-phase flow process is closely related to the sediment scouring around the monopile, which in turn affects the operation of the wind turbine structure. Therefore, research on this type of sediment carrying flow field has certain practical significance.

Prior research primarily focused on the flow around a fixed monopile. However, during operation, a monopile is susceptible to complex influences such as wind, waves, ocean currents, and the effect of pile-top machinery. These elements often demonstrate persistent and cyclical attributes that, coupled with soil stress, could potentially induce a monopile's cyclical vibration\cite{ref-3,ref-4,ref-5,ref-6,ref-7}. When such a vibration ensues, it disrupts the flow field around the pile, resulting in different morphological traits compared to those of a fixed monopile's flow. For instance, the shed vortex on the monopile's surface could reattach due to inertia, and the sheer zone near the base of the pile might undergo cyclical changes\cite{ref-9,ref-10,ref-11}.

The flow field around the pile, especially the eddy current development, often significantly influences sediment particles' movement in the sediment-laden flow\cite{ref-19,ref-20}. The sediment incipience Shields number usually characterizes the probability of particles being lifted and transported by the water flow from the bed surface. The force distribution on the sediment particles on the bed sand surface is relatively complex: the incipience of fine sediment particles typically requires considering factors such as sediment gravity, buoyancy, drag force, Saffman's lift force, particle cohesive force, and additional static water pressure. These factors correspond with both the sediment properties and the surrounding flow field\cite{ref-41,ref-40}. For instance, the bed sand's shielding of particles, sediment shape influence on drag force, wall boundary layer's effect on Saffman's lift force\cite{ref-45}, and particle spacing's impact on viscous particle cohesive force are all responsive to the flow field around the sediment. These combined effects result in varying incipience characteristics for different sediment particles in the pile's surrounding flow field at different times\cite{ref-42,ref-43}, subsequently affecting the scouring morphology around the pile.

Given the complexity of the water sediment two-phase flow, numerical simulation techniques have become increasingly crucial in such research\cite{ref-22,ref-23,ref-24}. High-precision simulation methods typically involve Direct Numerical Simulation (DNS) methods and Large Eddy Simulation (LES) methods. However, employing DNS for comprehensive flow field analysis necessitates substantial computational resources. Moreover, in this study, the particle size is smaller than the mesh size, hindering full-flow field analysis. Therefore, implementing the LES method for simulations provides not only high computational accuracy but also economizes computational resources by modeling small eddies instead of performing a full analysis\cite{ref-25,ref-26,ref-27}.

In summary, the research on the flow of sand laden water around vibrating piles comes from the practical background, and has certain practical significance and reference value for the operation and maintenance of offshore monopiles. In addition, the force on the particles near the wall is very complex, and the movement of these particles in that region is also worth considering. In Section 1 of this study, the relevant background of sediment initiation and particle flow simulation was introduced. In Section 2, combined with sediment initiation theory, the turbulence model was introduced and the momentum model of sediment particles on the bed was supplemented. A numerical water flume was constructed for research. In Section 3, the mesh of the monopile turbulent flow model was tested, and the sediment incipience model proposed in this study was validated based on classic research results. In Section 4, based on the calculation results, the influence of the flow field under different monopile vibration cases on the sediment incipient Shield number was compared and discussed. Finally, the results are summarized in Section 5.

\section{Method}

The Euler Lagrangian method was applied in this study for fluid and particle simulations. The fluid phase is regarded as a continuous phase and the filtered Navistok equation is solved on a small scale, while the particles are tracked using the Lagrangian method as discrete phases. In this study, the continuous phase (water) velocity is $\vec{u}=ux\vec{i}+vy\vec{j}+wz\vec{k}=u_{1}\vec{i}+u_{2}\vec{j}+u_{3}\vec{k}$, the discrete phase (particle) velocity is $\vec{u_{p}}=ux_{p}\vec{i}+vy_{p}\vec{j}+wz_{p}\vec{k}=u_{p1}\vec{i}+u_{p2}\vec{j}+u_{p3}\vec{k}$.

\subsection{Continuous phase model}
\subsubsection{Open channel flow model}

The incipience and scouring of sediment often occur at the bottom of open channel channels, so a water air two-phase fluid volume fraction model is used to simulate the actual environment of open channel flow:

\begin{subequations}
	\setlength{\abovedisplayskip}{1pt}
	\setlength{\belowdisplayskip}{1pt} 
	\begin{align}
		&\frac{\partial}{\partial t}\left(\alpha_{q} \rho_{q}\right)+\nabla \cdot\left(\alpha_{q} \rho_{q} \vec{v}_{q}\right)=S_{\alpha_{q}}+\sum_{p=1}^{n}\left(m_{p q}-m_{q p}\right) \\  \vspace{1ex}
		&\frac{\partial}{\partial t}(\rho \vec{u})+\nabla \cdot(\rho \vec{u} \vec{u})=-\nabla p+\nabla \cdot\left[\mu\left(\nabla \vec{u}+\nabla \vec{u}^{T}\right)\right]+\rho \vec{g} \\ \vspace{1ex}
	\end{align}
\end{subequations}

where 

\begin{subequations}
	\setlength{\abovedisplayskip}{1pt}
	\setlength{\belowdisplayskip}{1pt}
	\begin{align}
		&\rho=\alpha_{q} \rho_{q}+\left(1-\alpha_{q}\right) \rho_{p}\\ \vspace{1ex}
		&\mu=\alpha_{q} \mu_{q}+\left(1-\alpha_{q}\right) \mu_{p}
	\end{align}
\end{subequations}

$\alpha_{q}$ is volume fraction, $p$ is the  pressure in excess of the static pressure $\vec{g}$ is gravity acceleration vector, $\vec{F}$ is  $S$ is the source term(this term is 0 in this study), $m_{pq}$ is the mass transfer from phase $p$ to phase $q$, $m_{qp}$ is the mass transfer from phase $q$ to phase $p$.

$\alpha_{q}=0$, The cell is empty of the $q$ fluid;

$\alpha_{q}=1$, The cell is full of the $q$ fluid;

$0<\alpha_{q}<1$, The cell contains the interface between the $q$ fluid and one or more other fluids.

It is worth noting that only one phase velocity $\vec{u}$ is solved, and the phase velocity obtained is shared by all phases.

\subsubsection{WALE model}

The vortex field around the pile has a significant impact on the bed velocity gradient, which in turn affects the initiation of sediment on the bed, and is one of the simulation focuses of this study. 

In this study, the flow variables $\phi$ are decomposed into a largescale component (denoted by an widetilde) and a small subgrid-scale component by applying a filtering operation:

\begin{equation}
	\phi(x)=\widetilde{\phi}(x)+\phi^{\prime}(x)
\end{equation}
where
\begin{equation}
	\widetilde{\phi}(x)=\int_{D} \phi\left(x\right) G\left(x, x^{\prime}\right) d x^{\prime}
\end{equation}
$x^{\prime}$ is the location coordinate of another fluid particle, $D$ is the fluid domain. The box filtering method was adopted in this study:

\begin{equation}
	\setlength{\abovedisplayskip}{1pt}
	\setlength{\belowdisplayskip}{1pt}
	G\left(x,\ x^{\prime}\right)=\left\{\begin{array}{l}
		1/V,|x-x^{\prime}|<{\Delta/2} \\
		0, \text\ {\ otherwise\ }
	\end{array}\right.
\end{equation}
where $x^{\prime}$ is the location coordinate of another fluid particle, $V$ is the volume of a computational cell, $\Delta=V^{1/3}$.

Filtering the continuity and momentum equations:

\begin{subequations}
	\setlength{\abovedisplayskip}{1pt}
	\setlength{\belowdisplayskip}{1pt}
	\begin{align}
		&\frac{\partial\ \rho}{\partial\ t}+\frac{\partial}{\partial\ x_{i}}\left(\rho\ \widetilde{u}_{i}\right)=0 \\
		&\frac{\partial}{\partial\ t}\left(\rho\ \widetilde{u}_{i}\right)+\frac{\partial}{\partial\ x_{j}}\left(\rho\ \widetilde{u}_{i}\ \widetilde{u}_{j}\right)=\frac{\partial}{\partial\ x_{j}}\left(\sigma_{ij}\right)-\frac{\partial\ \widetilde{p}}{\partial\ x_{i}}-\frac{\partial\ \tau_{i\ j}}{\partial\ x_{j}}+F_{i}\\
	\end{align}
\end{subequations}

The first equation is continuity equation, and the terms of the second equation are Transient term, Convective term, Stress tensor term, Pressure term, Subgrid scale stress term and Particle force, respectively.

where $\rho$ is the density of continuous phase,

\begin{equation}
	F_{i}=-\left.\frac{1}{V} \sum_{n_{i}=1}^{N} (\vec{F_{D p}}+\vec{F_{gp}}+\vec{F_{vir}}+\vec{F_{pre}}+\vec{F_{sf}})\right|_{n_{i}}
\end{equation}

$\vec{F_{D p}}$ means the drag force, $\vec{F_{g p}}$ means the gravity and buoyancy on particle,  $\vec{F_{vir}}$ means the virtual mass force, $\vec{F_{pre}}$ means the pressure gradient force, $\vec{F_{sf}}$ means the Saffman's lift(for details in \ref{secpartf}), $N$ is the number of particles in single grid.

$\sigma_{i j}$ is the molecular viscosity stress tensor:
\begin{equation}
	\sigma_{i j} =\left[\mu\left(\frac{\partial \widetilde{u}_{i}}{\partial x_{j}}+\frac{\partial \widetilde{u}_{j}}{\partial x_{i}}\right)\right]-\frac{2}{3} \mu \frac{\partial \widetilde{u}_{l}}{\partial x_{l}} \delta_{i j}
\end{equation}
$\mu$ means the dynamic viscosity of fluid(water,$\mu=0.001003 (kg\cdot m^{-1}\cdot s^{-1})$). $\delta_{i j}$ is the unit tensor.
$\tau_{i j}$ is the subgrid scale stress:
\begin{equation}
	\tau_{i j} = \rho \widetilde{u_{i} u_{j}}-\rho \widetilde{u}_{i} \widetilde{u}_{j}
\end{equation}
$\mu_{td}$ is eddy viscosity in LES model:
\begin{equation}\label{equ mu t}
	\mu_{td}=\rho L_{s}^{2} \frac{\left(S_{i j}^{d} S_{i j}^{d}\right)^{3 / 2}}{\left(\widetilde{S}_{i j} \widetilde{S}_{i j}\right)^{5 / 2}+\left(S_{i j}^{d} S_{i j}^{d}\right)^{5 / 4}}
\end{equation}
where 
\begin{subequations}\label{equ ls}
	\setlength{\abovedisplayskip}{1pt}
	\setlength{\belowdisplayskip}{1pt}
	\begin{align}
		&\widetilde{S}_{i j} = \frac{1}{2}\left(\frac{\partial \widetilde{u}_{i}}{\partial x_{j}}+\frac{\partial \widetilde{u}_{j}}{\partial x_{i}}\right)\\
		&L_{s}=\min \left(\kappa d, C_{w} V^{1 / 3}\right) \\
		&S_{i j}^{d}=\frac{1}{2}\left(\widetilde{g}_{i j}^{2}+\widetilde{g}_{j i}^{2}\right)-\frac{1}{3} \delta_{i j} \widetilde{g}_{k k}^{2} \\
		&\widetilde{g}_{i j}=\frac{\partial \widetilde{u}_{i}}{\partial x_{j}}
	\end{align}
\end{subequations}
$\kappa=0.41$, $C_{w}=0.325$.

The advantage of above model is that it can simulate possible laminar flow situations without causing additional dissipation.

\subsection{Discrete Phase Model}\label{sec2.2}

The continuous flow field was obtained through fluid calculation in Section 2.1, and then the calculated flow field was used to track a large number of particles in the flow field to solve the problem of dispersed phase distribution. The dispersed phase can exchange momentum, mass, and energy with the continuous phase (mainly momentum exchange in this study).

\subsubsection{Particle Momentum Model}\label{secpartf}

The force balance equation on the particle, which is written in a Lagrangian reference frame is:
\begin{equation}
	m_{p} \frac{d \vec{u}_{p}}{d t}=m_{p} \frac{\vec{u}-\vec{u}_{p}}{\tau_{r}}+m_{p} \frac{\vec{g}\left(\rho_{p}-\rho\right)}{\rho_{p}}+\vec{F_{others}}
\end{equation}

$\rho_{p}$ is the density of discrete phase, $m_{p}$ means the mass of single particle,The first term on the right of the equation is \textbf{Drag force}, where
\begin{equation}
	{\tau_{r}}=\frac{\rho_{p} d_{p}^{2}}{\mu} \frac{24}{18 C_{D} Re_p}
\end{equation}
$d_{p}$ is particle diameter,$C_{D}$ is the drag coefficient. For spherical particles in this study:
\begin{equation}
	C_{D}=a_{1}+\frac{a_{2}}{R e_{p}}+\frac{a_{3}}{R e_{p}^{2}}
\end{equation}
$Re_{p}$ is the Reynold number of particle: 
\begin{equation}\label{equ rep}
	Re_{p}={\rho}|u_p-u|d_p/{\mu}
\end{equation}
$a_1$, $a_2$ and $a_3$ are the quantity varying with $Re_{p}$\cite{ref-29}: For $Re_{p}$ <0.1, $a_1$=0,$a_2$=24, $a_3$=0; For 0.1<$Re_{p}$ <1.0, $a_1$=3.69, $a_2$=22.73, $a_3$=0.0903; For 1.0< $Re_{p}$ <10.0, $a_1$=1.222, $a_2$=29.1667, $a_3$=-3.8889\cite{ref-29}.

$\vec{F_{others}}$ include:

\textbf{virtual mass force}:The force required to accelerate the fluid surrounding the particle due to the difference in acceleration between particles and fluids:
\begin{equation}
	\vec{F_{vir}}=C_{v m} m_{p} \frac{\rho}{\rho_{p}}\left(\vec{u}_{p} \nabla \vec{u}-\frac{d \vec{u}_{p}}{d t}\right)
\end{equation}
where $C_{vm}$ is the virtual mass factor with a default value of 0.5.

\textbf{pressure gradient force}: Another force arises due to the pressure gradient:
\begin{equation}
	\vec{F_{pre}}=m_{p} \frac{\rho}{\rho_{p}} \vec{u} \nabla \vec{u}
\end{equation}

Within the boundary layer, the pressure gradient in the $Y$-direction varies less, so the pressure gradient forces in the $X$ and $Z$ directions are mainly considered.

\subsection{Overall algorithm}
The pseudo code \ref{alg1} show the overall algorithm of coupling model. The continuous phase equation and the discrete phase equation are computed step by step and coupled through particle interaction forces.

\begin{algorithm}[H]
	\SetAlgoLined
	\caption{The overall algorithm of coupling model}
	\label{alg1}
	\SetKwInOut{Input}{Input}
	\SetKwInOut{Output}{Output}
	\Input{Initial conditions for continuous and discrete phase}
	\Output{Final state of the continuous and discrete phase}
	\BlankLine
	\SetKwData{velocity}{velocity}
	\SetKwData{pressure}{pressure}
	\SetKwData{discreteVelocity}{discreteVelocity}
	\SetKwData{position}{position}
	\SetKwData{force}{force}
	\SetKwData{shieldsNumber}{shieldsNumber}
	\BlankLine
	\velocity $u$ $\leftarrow$ InitializeContinuousVelocity\;
	\pressure $p$$\leftarrow$ InitializePressure\;
	\discreteVelocity $u_p$ $\leftarrow$ InitializeDiscreteVelocity\;
	\position $x_p$ $\leftarrow$ InitializeDiscretePosition\;
	\position $F_i$ $\leftarrow$ InitializeParticleForce\;
	\BlankLine
	\SetKwFunction{solveContinuityEquation}{SolveContinuityEquation}
	\SetKwFunction{solveMomentumEquationContinuous}{SolveMomentumEquationContinuous}
	\SetKwFunction{solveMomentumEquationDiscrete}{SolveMomentumEquationDiscrete}
	\BlankLine
	\While{$t < T_{end}$}{
		$ u' \leftarrow$ \solveContinuityEquation{$ u$}\;
		$u', p' \leftarrow$ \solveMomentumEquationContinuous{$u', p, F_{i}$}\;
		$d_{p}', x_{p}'$, ParticleForce$F_{i}$, ShieldsNumber$\theta_{ci}$ $\leftarrow$ \solveMomentumEquationDiscrete{$d_{p}, x_{p}$}\;
		\BlankLine
		$t \leftarrow t + dt$\;
	}
	\BlankLine
	Output $u$ (Continuous Phase Velocity), $p$ (Continuous Phase Pressure), $d_{p}$ (Discrete Phase Velocity), $x_{p}$ (Discrete Phase Position), $\theta_{ci}$ (Particle Shields numbers)
	\caption{The overall algorithm of coupling model}
\end{algorithm}

The SIMPLEC algorithm was applied in this study. The discretization schemes of terms are:
\textbf{Continuity equation}:The second order central differencing scheme;

\textbf{Transient term}:The second order central differencing scheme;

\textbf{Convective term}:The QUICK scheme;

\textbf{Stress tensor term:}:The second order central differencing scheme;

\textbf{Subgrid-scale stress term}:The QUICK scheme;

\textbf{Pressure term}:The modified body force weighted scheme;

\textbf{Particle motion term}:The third order central differencing scheme.

\subsection{Boundary conditions}
\subsubsection{Continuous phase boundary conditions}

\textbf{Inlet}: Open channel inlet with $H=0.1 m$. Inlet velocity is obtained by giving velocity components(for details in subsection \ref{2.5.2});

\textbf{Outlet}:Open channel outlet with $H=0.1 m$.

\textbf{Other walls}:The side walls and the top wall of flume are symmetry boundaries;  $\Delta_p=2.5 d_p$

\subsubsection{Discrete phase boundary conditions}

\textbf{Inlet and outlet}:Escape boundaries.

\textbf{Other walls}:The discrete phase boundary of bed surface and the discrete phase boundary of pile sidewall is reflecting boundary.  The following empirical correlations are used for the collision restitution ratio:\cite{ref-17}:

\begin{subequations}
	\setlength{\abovedisplayskip}{1pt}
	\setlength{\belowdisplayskip}{1pt}
	\begin{align}
		&\frac{V_{n_{2}}}{V_{n_{1}}}=1.0-0.4159 \beta-0.4994 \beta^{2}+0.292 \beta^{3} \\
		&\frac{V_{t_{2}}}{V_{t_{1}}}=1.0-2.12 \beta+3.0775 \beta^{2}-1.1 \beta^{3}
	\end{align}
\end{subequations}

$V_{ni}$ and $V_{ti}$ represent the particle collision velocity components normal and tangential to the wall before and after collision, respectively, $\beta$ is the coefficient of restitution which represents the angle between the incident velocity and the tangent to the surface.

\subsection{Vortex identification}

The vortex structure around the pile has a significant impact on the convective field, which in turn affects the movement of sediment around the pile. Therefore, the correct identification of vortices around piles is crucial for the analysis of the results in this study. The Liutex vortex identification method was adopted in this study\cite{ref-31, ref-32}:

\begin{equation}
	\setlength{\abovedisplayskip}{1pt}
	\setlength{\belowdisplayskip}{1pt}
	\omega_L=R+S
\end{equation}

where $R$ represents the vorticity of the rotating part and $S$ denotes the vorticity of the non-rotating part. Subsequently, a numerical value $\Omega_L$ is introduced to represent the proportion of the rotating vorticity to the total vorticity \cite{ref-27,ref-28}:

\begin{equation} \label{eqomg}
	\setlength{\abovedisplayskip}{1pt}
	\setlength{\belowdisplayskip}{1pt}
	\Omega_L=\frac{\|B\|_{F}^{2}}{\|A\|_{F}^{2}+\|B\|_{F}^{2}+\varepsilon}
\end{equation}

where $\varepsilon=0.002$,

\begin{equation}
	\setlength{\abovedisplayskip}{1pt}
	\setlength{\belowdisplayskip}{1pt}
	A=\left[\begin{array}{ccc}
		\frac{\partial u_x}{\partial x} & \frac{1}{2}\left(\frac{\partial u_x}{\partial y}+\frac{\partial u_y}{\partial x}\right) & \frac{1}{2}\left(\frac{\partial u_x}{\partial z}+\frac{\partial u_z}{\partial x}\right) \\
		\frac{1}{2}\left(\frac{\partial u_y}{\partial x}+\frac{\partial u_x}{\partial y}\right) & \frac{\partial u_y}{\partial y} & \frac{1}{2}\left(\frac{\partial u_y}{\partial z}+\frac{\partial u_z}{\partial y}\right) \\
		\frac{1}{2}\left(\frac{\partial u_z}{\partial x}+\frac{\partial u_x}{\partial z}\right) & \frac{1}{2}\left(\frac{\partial u_z}{\partial y}+\frac{\partial u_y}{\partial z}\right) & \frac{\partial u_z}{\partial z}\\
		
	\end{array}\right]
\end{equation}

\begin{equation}
	\setlength{\abovedisplayskip}{1pt}
	\setlength{\belowdisplayskip}{1pt}
	B=\left[\begin{array}{ccc}
		0 & \frac{1}{2}\left(\frac{\partial u_x}{\partial y}-\frac{\partial u_y}{\partial x}\right) & \frac{1}{2}\left(\frac{\partial u_x}{\partial z}-\frac{\partial u_z}{\partial x}\right) \\
		-\frac{1}{2}\left(\frac{\partial u_x}{\partial y}-\frac{\partial u_y}{\partial x}\right) & 0 & \frac{1}{2}\left(\frac{\partial u_y}{\partial z}-\frac{\partial u_z}{\partial y}\right) \\
		-\frac{1}{2}\left(\frac{\partial u_x}{\partial z}-\frac{\partial u_z}{\partial x}\right) & -\frac{1}{2}\left(\frac{\partial u_y}{\partial z}-\frac{\partial u_z}{\partial y}\right) & 0
	\end{array}\right]
\end{equation}

$ \Omega_L \in (0,1)$. Obviously, when the value is higher than 0.5, the rotating part at this location dominates, and it can be considered that vortices are generated here.

\subsection{Numerical flume model}
\subsubsection{Numerical flume}
During the operation of monopile, a monopile is susceptible to complex influences such as wind, waves, ocean currents, and the effect of pile-top machinery. These elements often demonstrate persistent and cyclical attributes that, coupled with soil stress, could potentially induce a monopile's cyclical vibration\cite{ref-3,ref-4,ref-5,ref-6,ref-7}. To facilitate the summary of the impact of monopile vibration on the flow field, in this study, the vibration period of the monopile was set to be consistent with the vortex shedding period.

The numerical flume for cylindrical flow (with cylindrical diameter of $D$). The model measures $50D$ in length, $10D$ in width, and $15D$ in height. A cylindrical monopile with a height of $15D$ is arranged strategically placed at the center of the flume( extending from the top to the bottom).

For convenience in the following description, a three-dimensional rectangular coordinate system is established with the centre point of the section of monopile with the calm water surface as the coordinate origin, the positive direction of wave propagation in the flume as the $X$-axis, the height of the flume (from bottom to top) as the $Y$-axis, and the width of the flume (from left to right facing the wave propagation direction) as the $Z$-axis.

Considering the flow field characteristics around the monopile in this study, the meshes on the side wall and bed surface are refined 

\subsubsection{Monopile vibration}\label{2.5.2}

During the operation of monopile, a monopile is susceptible to complex influences such as wind, waves, ocean currents, and the effect of pile-top machinery. These elements often demonstrate persistent and cyclical attributes that, coupled with soil stress, could potentially induce a monopile's cyclical vibration\cite{ref-3,ref-4,ref-5,ref-6,ref-7}. 

Due to the use of box filtering method in this study, the turbulent subgrid model is closely related to the mesh size, so it is not possible to use dynamic mesh method to simulate monopile vibration. Therefore, relative water flow velocity is set at the inlet to simulate the situation of monopile vibration:

\begin{equation} \label{uinlet}
	\setlength{\abovedisplayskip}{1pt}
	\setlength{\belowdisplayskip}{1pt}
	U_{inlet}=U_{0}+U_{A}sin(\omega_{p}(t-t_0))
\end{equation}

$U_0$ is the flow rate, $U_{A}$is the amplitude of monopile vibration velocity variation,$t_{0}$ is the initial time when monopile begins to vibrate (in this study, the moment of complete detachment of the left vortex during the vortex shedding period is taken as this initial time), $\omega_{p}=\frac{2 \pi}{T_{p}}$, $T_{p}$ is the vibration period of the monopile. To facilitate the summary of the impact of monopile vibration on the flow field, in this study, the vibration period of the monopile was set to be consistent with the vortex shedding period(for details in Table \ref{tab5}).

\section{ Validation of numerical model}
\subsection{ Validation cases}
As shown in Table \ref{tab1}, two validation cases were set up in the numerical water flume to validate the turbulent flow model around monopile and the sediment incipience model:

\begin{table*}
	\caption{\label{tab1}Validation cases(a).}
	\begin{ruledtabular}
		\begin{tabular}{ccccc}
			\textbf{Index}& {flow rate($m/s$)}&{monopile diameter($m$)}& {particle diameter($\mu m$)} & {particle density($kg/m^{3}$)}\\
			\hline
			Case(a)  	& 0.1	   &0.01         & - & - \\
			Case(b) 	& $U_{b}(t)$	   &0.01         & $d_{b}i$ & 2650
		\end{tabular}
	\end{ruledtabular}
\end{table*}

\begin{subequations}
	\setlength{\abovedisplayskip}{1pt}
	\setlength{\belowdisplayskip}{1pt}
	\begin{align}
		U_{b}(t)=\left\{\begin{array}{c}
			0.06,0<t<3 \\
			0.06+0.02(t-3), t \geq 3
		\end{array}\right.
	\end{align}
\end{subequations}

$d_{b}i = 5,10,15,50,100,200,400$

\subsection{ Comparison with previous flow around circular cylinder studies}

Compare the calculation results with previous research results on the flow around cylinder\cite{ref-46}(as shown in Table \ref{tab2}):

\begin{table}
	\caption{\label{tab2}Comparison between Case(a) and classical results}
	\begin{ruledtabular}
		\begin{tabular}{cccc}
			\textbf{Index}	&{$Re$}	    &{$Cd_{ave}$}  & {$St$}\\
			\hline
			Case(a)  	& 1000	   & 1.15         & 0.23    \\
			Cao Shuyang & 1000	   &1.10         & 0.22 \\ 
			Tamura   	& 1000	   &1.02        & 0.22 \\ 
		\end{tabular}
	\end{ruledtabular}
\end{table}

Where $Cd_{ave}$ is the average count of drag coefficient, $St$ is the Strouhal number, $St=\frac{f_{p} D}{U_{inlet}}$, $f_{p}$ is the vortex shedding frequency. These findings align with previous research results.

\subsection{Verification of particle effect on velocity profile}
The velocity profile directly influences vortex shedding, making it the most direct factor in this study for affecting particle distribution through the flow field. Compare the particle effect on velocity profile near the side wall with Yohei Sato's classic research\cite{ref-35} 

The above results show that the particle distribution pattern obtained in this study has credibility.

\subsection{ Comparison with classical sediment incipience studies}
In Case (b), 150 particles are aligned side by side at the bottom surface of the inlet section of the water flume, with an even distribution. The water phase velocity at the flume inlet is then gradually increased while the particle incipience is observed and the average velocity on the velocity profile where the activated particles reside is recorded. The recorded results are then compared with the formulas proposed by Dou Guoren and Zhang Ruijin.\cite{ref-40}:

Dou Guoren formula:

\begin{equation}
	\begin{split}
		ux_s=2.5 ln(11 \frac{H}{\Delta_p}) 1.75 (\frac{\Delta_p}{0.01})^{0.167} (58.2512 d_p +\\ \frac{0.00000175}{d_p}+\frac{0.0000022653 H (0.000000231/d_p)^{0.5}}{d_p})^{0.5}
	\end{split}
\end{equation}

Zhang Ruijing formula:

\begin{equation}
	ux_s=(\frac{H}{d_p})^{0.14} (29 d_p +0.000000605 \frac{10+H}{d^{0.72}})^{0.5}
\end{equation}

It can be observed that the majority of the calculated results are greater than those obtained using Dou Guoren's formula, yet smaller than those from Zhang Ruijin's formula. This discrepancy is attributed to variations in the computation of sediment drag force compared with the empirical formulas. Examining the calculated outcomes reveals that the incipient flow rate is at its lowest point when about $d_{b}i=100\mu m$. Both larger and smaller particle sizes lead to an increase in the incipient flow rate. Larger particles possess greater weight, making it more challenging for them to move due to gravity. Conversely, smaller particles are chiefly influenced by particle cohesive force and supplemental static water pressure, rendering them less prone to motion. These findings align with previous research results.

\section{Result and Discussion}

Four cases with different monopile amplitudes were set in this study as shown in Tab.\ref{tab5}:
\begin{table*}
	\caption{\label{tab5}Working cases.}
	\begin{ruledtabular}
		\begin{tabular}{ccccccc}
			\textbf{Index}&{$U_{0}$($m/s$)}&{particle diameter($\mu m$)}&{particle density($kg/m^{3}$)}&{pile amplitude($mm$)} &{vibration period($s$)}\\
			\hline
			Case1  	& 0.15	   &50         & 2650   &0   &0.298 \\
			Case2  	& 0.15	   &50         & 2650  &3.766   &0.298\\ 
			Case3  	& 0.15	   &50         & 2650  &7.531   &0.298\\
			Case4  	& 0.15	   &50         & 2650  &11.297   &0.298\\
		\end{tabular}
	\end{ruledtabular}
\end{table*}
According to the previous trial calculations, it is obtained that the vortex shedding period is 0.298 $s$ in this study. To examine the circumstances of particle incipience, depicted , 6400 particles are positioned around the pile. Each particle center aligns with the center of the grid and is arranged in correspondence with the mesh density, thus yielding a higher particle density in the vicinity of the monopile.

Given that the initial velocity of the particles is set at (0,0,0), as indicated in Chapter \ref{sec2.2.3}, it would be pragmatic to consider that the larger the absolute value of the particle's improved Shields number, the greater likelihood for these particles to be eroded in the corresponding direction.

\subsection{Distribution Characteristics of particle improved Shields Number }
 The bed surface region around the monopile can be segmented into three distinct zones: the deceleration region situated in front of the pile, the shear acceleration region located on the side of the pile, and the wake region at the rear of the pile. The characteristics of the particle incipience's improved Shields number vary across the different regions. In this study, the specific area's average improved Shields number for particle incipience, $Ave_ {uip}$, and the dispersion function, $Std_ {uip}$, are used to characterize the regional particle incipience characteristics:

\begin{subequations}
	\setlength{\abovedisplayskip}{1pt}
	\setlength{\belowdisplayskip}{1pt}
	\begin{align}
		&Ave_{uip}=\frac{\sum\limits_{p=1}^{n_p} \theta_{cij}}{n_{p}}\\\vspace{1ex}   
		&Std_{uip}=(\frac{(\theta_{cij}-Ave_{uip})^{2}}{n_{p}})^{0.5}
	\end{align}
\end{subequations}

where $n_p$ is the number of particles in the region,$i={x,z}$ represents the coordinate, direction,$j={1,2,3,......,n_p}$ represents the count number of particles.

\subsubsection{The deceleration region situated in front of the pile}
Within the deceleration zone positioned in front of the pile, the reduced flow velocity leads to a corresponding decrease in the particle's bottom flow velocity, making particle incipience less likely to occur. Consequently, the count of improved Shields numbers for particle incipience remains relatively low. As represented , a few incidences of reverse particle incipience are observed in an arc-like pattern immediately preceding the pile's base. This outcome can be attributed to the horseshoe vortices existing in front of the pile; this vortex-induced backflow creates a dredging effect on the particles ahead of the pile, and these findings align with the previous research results.

When comparing the regions in front of the piles in Case1$\sim$4, it's evident that the vibrations of the monopile create an additional reflux region both in front of and behind the monopile. Due to this effect, the intensity of the horseshoe vortex in front of the pile heightens, thereby amplifying the dredging impact. The consequences faced by the particles manifest as an escalated number of reverse direction particle incipience within the region ahead of the pile, as well as an increased count of improved Shields numbers for particle incipience in the negative $X$ direction, which is clearly depicted . On a similar scale, it can be observed that the reverse particle incipience area becomes distinctly visible as the amplitude increases. This characteristic can be revealed in a short duration (within $ \omega t \in (0^ {\circ},60^ {\circ}) $) following the initiation of vibration.

Given the symmetry on the $X$-axis within this region, the number of improved Shields for particle movement in the $Z$ direction is also noteworthy. With its symmetry taken into account, the dispersion function value of $Z$-direction velocity is employed to represent the statistical features of the improved Shields number in the $Z$ direction. As conveyed in Figure \ref{fig4.4}, the particles exhibit a propensity for bi-directional dispersion along the z-axis. Simultaneously, it's noticeable that the greater the amplitude, the more pronounced the lateral diffusion of the initial particle flow rate is, especially in Case 4. The dispersion function time history of the improved Shields number peaks within the range of $ \omega t \in (200^ {\circ}, 250^ {\circ}) $, and subsequently diminishes. This trend aligns with the lateral expansion of the reflux region in front of the pile.

\subsubsection{The shear acceleration region located on the side of the pile}\label{sec4.2.2}

Within the shear acceleration region on the side of the pile, the intensity of the shear flow hinges directly on the relative flow velocity between the single pile and the inlet. Upon examining Cases 1 $\sim $4 in Figure \ref{fig4.6}, it becomes evident that the positioning of the high shear zone fluctuates with the monopile's vibration period. In tandem, Figure \ref{fig4.17} represents a similar "oscillation" in the distribution of improved Shields numbers for particle incipience. The magnitude of the monopile's amplitude directly influences the conspicuousness of this "swinging" pattern. Intriguingly, the average count of improved Shields numbers for particles presents a swinging pattern that first dips, then rises, and ultimately falls again, which doesn't completely conform to the alterations in the monopile's relative velocity compared to the incoming flow in this study. This discrepancy corroborates the impact of detachment vortices on the region. The monopile's vibration initiates at the precise moment when the right vortex completely sheds off, and this includes the processes of the left vortex shedding ($\omega t \in (0^ {\circ},180^ {\circ}) $) as well as the reformation of the right vortex ($\omega t \in (180^ {\circ},360^ {\circ}) $). 

Correspondingly, on the left side of the monopile, the peak of the average improved Shields number is conspicuously presented near $\omega t=180^ {\circ} $. That is, under the influence of the left-side shedding vortex, the average improved Shields number for the vortex-affected particles persistently increases up to $ \omega t=180^ {\circ} $. Meanwhile, on the right side of the pile, within the timeframe of $\omega t \in (0^ {\circ},150^ {\circ}) $, the complete detachment of the right vortex leads to a gradual loss of driving force for the particles, and a continuous decline in the count of the average improved Shields number. Not until the point of $ \omega t=180^ {\circ} $, where the right vortex reforms, do the particles regain their driving force and the average count of the improved Shields number escalates yet again. Contrary to the trend of the pile's linear velocity, the trend of the average count of the improved Shields number for the particles on the right side indicates that the detachment vortices are the predominate driving factors for the particles within this region.

Upon comparison of the average count of improved Shields in Case1$\sim$4, it becomes evident that a larger amplitude of the pile corresponds to a greater count of the average count of the  improved Shields. Within the time span of $ \omega t \in (0^ {\circ},180^ {\circ}) $, the monopile is within the staring cycle, and the relative flow velocity escalates in line with the incoming flow. Consequently, the adverse pressure gradient on the side wall of the monopile intensifies, permitting the right vortex to garner more vorticity from the side wall. This study employs equation ref to portray the average intensity of the shedding vortices:

A larger amplitude leads to a heightened intensity of the right vortex, thereby endowing the bottom particles with a more potent driving force. As demonstrated by the time history curve of vortex intensity, a larger amplitude corresponds to a higher vortex intensity. This is attributable to the pile's vibration amplifying the vorticity on its surface, allowing the shedding vortex to acquire a more potent intensity from the surface, thereby aligning with an escalation in the figure's flow velocity. Another remarkable phenomenon is the secondary peak of vortex intensity noticeable on the right side of both Case3 and Case4. This suggests that, within the context of this study, a larger monopile amplitude contracts the period of vortex shedding. Under the drive of the shedding vortex, the count of the improved Shields for particles in the right region reciprocally adjusts.

\subsubsection{The wake region at the rear of the pile}
In the wake region, it is noteworthy that a persistently evolving negative $X$ incipience region for particles exists. An examination of the vortex distribution behind the pile reveals the presence of a recirculation zone nestled between the shedding vortices. This reflux region confines the particles and triggers a reverse incipience phenomenon.

Upon examining Case1 $\sim $4, it becomes evident that the positioning of the reverse incipience area remains largely the same across different Cases. However, due to the varying amplitudes of the monopiles, the count of the improved Shields for particles differs. The statistics depicted in Figure \ref{fig4.10} indicate that a higher amplitude corresponds to a greater count of improved Shields in the negative $X$ direction, a pattern maintained throughout the entire cycle. Thus, it becomes apparent that an increased amplitude elevates the likelihood of particle movement towards the negative $X$ direction. It is also noteworthy that a larger amplitude causes a more sizable fluctuation in the time history curve of the average improved Shields count, an observation aligning with the continuous expansion and contraction of the reverse incipience region in Figure \ref{fig4.18}. Interestingly, these fluctuations are more conspicuous as the amplitude escalates. Importantly, within the time span of $ \omega t \in (180^ {\circ}, 240^ {\circ}) $ during the latter half of the cycle, the vibration line velocity of the monopile declines, yet a peak becomes apparent in the time history curve of the negative $X$ average improved Shields count. This accentuates the impact of detachment vortices on the reflux region. Based on the analysis of \ref{sec4.2.2}, during this timeline, the right vortex recovers its momentum, and the main vortex separating from the vortex on the right establishes a recirculation zone in the wake region, instigating a negative X direction incipience of the particles.

Considering the symmetry of this region, it's significant to note the improved Shield count of particles commencing in the $Z$ direction. As illustrated by the average count of the improved Shield number, the dispersion of the lateral improved Shield number in this area exhibits negligible variance among the four cases compared to other statistics. During the vibration period of monopile, the dispersion function continuously oscillates, mirroring the "oscillatory" characteristics of particles in the reflux region, which can also be inferred from the flow velocity distribution . Although the recirculation zone is instigated by the main vortex of the shedding vortex, it is profoundly influenced by the monopile's vibrations. Following the moment of $ omega t=180 ^ {\circ} $, the linear velocity of the monopile's vibrations decreases until it halts. Consequently, the velocity gradient between the detached vortex and the region elevates, enhancing the intensity of the reflux region. This also elucidates why the negative $X$ of particles in this region continues to escalate towards the average improved Shield count during this time frame, leading to amplified differences between various working conditions.

\section{Conclusions}
This study introduced a discrete phase momentum model for the unique stress conditions of bed particles and utilized this model to simulate the particle incipience around the vibrating monopile. The primary conclusions are as follows: 

(1)The coupling model utilized in this study successfully replicates the vortex and particle motion around the pile by modifying the effects of Saffman's lift. The model exhibits considerable precision in depicting the intricacies of vortex separation and reattachment, as well as the dispersion of particles.

(2) The particles preceding the pile are impacted by the horseshoe vortex, presenting a negative flow incipience phenomenon, with a predisposition to spread towards both sides of the pile; 

(3) The particle incipience trend on the pile's side is strongly influenced by the shedding vortex. In turn, the robust shedding vortex induced by the pile's larger amplitude in the concurrent period often results in a higher particle improvement Shields number in the region, fostering a more probable occurrence of the particle incipience phenomenon; 

(4) The particles in the wake reflux area are impacted by both shedding vortices and pile vibration. During a monopile's stopping cycle, the wake reflux area's reflux intensity escalates, leading to an increased negative flow direction of particles and an improved shield count. Consequently, the negative particle incipience phenomenon has a higher likelihood of taking place;

\section{Credit authorship contribution statement}
\textbf{Yuxuan Shi}: Validation, Data Curation, Writing - Original Draft; \textbf{Yongzhou Cheng}: Conception, Funding acquisition, Supervision, Writing - Review \& Editing.

\section{Declaration of competing interest}
The authors declare that they have no known competing financial  interests or personal relationships that could have appeared to influence  the work reported in this paper.

\section{Acknowledgements} 
This study was supported by the National Natural Science Foundation of China (Grant No.52071031; 52371258), the Open Foundation of State Key Laboratory of Coastal and Offshore Engineering, Dalian University of Technology (Grant No. LP21V3), Postgraduate Research and Innovation Project of Changsha University of Science \& Technology (Grant No. CX20210782).

\end{document}